# The Underluminous Nature of Sgr A*

F. Yusef-Zadeh* and M. Wardle†

*Dept Physics and Astronomy, Northwestern University, Evanston, IL. 60208, USA
†Department of Physics and Engineering, Macquarie University, Sydney NSW 2109, Australia

**Abstract.**
In the last several years, a number of observing campaigns of the massive black hole Sgr A* has been carried out in order to address two important issues: one concerns the underluminous nature of Sgr A* with its bolometric luminosity being several orders of magnitude less than those of its more massive counterparts. It turns out that the angular momentum of the ionized stellar winds from orbiting stars in one or two disks orbiting Sgr A* could be a critical factor in estimating accurately the accretion rate unto Sgr A*. A net angular momentum of ionized gas feeding Sgr A* could lower the Bondi rate. Furthermore, the recent time delay picture of the peak flare emission can be understood in the context of adiabatic expansion of hot plasma. The expansion speed of the plasma is estimated to be sub-relativistic. However, relativistic bulk motion of the plasma could lead to outflow from Sgr A*. Significant outflow from Sgr A* could then act as a feedback which could then reduce Bondi accretion rate. These uncertain factors can in part explain the underluminous nature of Sgr A*. The other issue is related to the emission mechanism and the cause of flare activity in different wavelength bands. The emission mechanism of flaring in the IR is now recognized to be due to synchrotron emission dominating the variable and quiescent emissions from Sgr A* in all wavelength bands with the possible exception of the X-ray emission. Modeling of X-ray and near-IR flares suggests that inverse Compton scattering (ICS) of IR flare photons by the energetic electrons responsible for the submm emission can account for for the X-ray flares. A time delay of minutes to tens of minutes is predicted between the peak flaring in the near-IR and X-rays, *not* due to adiabatic expansion of optically thick hot plasma, but to the time taken for IR flare photons to cross the accretion flow before being upscattered. Indeed, the observed X-ray fluxes and lack of detected time delays place significant restrictions on the electron density and temperature profile in the inner accretion flow. These constraints are tightened if the X-ray flares are instead produced by other mechanisms, such as synchrotron emission or the synchrotron self-Compton process as then the inverse Compton process can only provide a small fraction of the observed X-ray flare emission.

**PACS:** 90

## ACCRETION LUMINOSITY

The luminosity of Sgr A* is thought to be due to partial capture of thermal winds from a neighboring cluster of massive stars. However, the bolometric luminosity of Sgr A* ($\sim 100~L_\odot$) is several orders of magnitude lower than the Eddington luminosity (Baganoff et al. 2003). To explain this puzzle, we need to consider two important parameters on which the accretion luminosity depend. One is the fraction of the ionized stellar winds that is captured in the outer disk of Sgr A* due to Bondi-Hoyle accretion. The other is the radiative efficiency of the accreting gas in the inner disk. Broadly speaking, accurate estimates of these two uncertain parameters should clarify the puzzling low luminosity nature of Sgr A*.

## Bondi Accretion Rate

The rate at which ionized stellar winds from massive stellar winds accrete onto the outer disk of Sgr A* was originally estimated from near-IR measurements of mass-losing, young, massive stars (Wardle and Yusef-Zadeh 1992; Melia 1992). A total mass loss-rate of $\sim 10^{-3}$ $M_\odot yr^{-1}$ was estimated from the cluster of massive stars surrounding Sgr A*. The Bondi-Hoyle accretion rate was estimated by using the total outflow rate from the cluster. A more accurate accretion rate was calculated by incorporating the mass-loss rate of individual stars at different locations (Coker & Melia 1997). This was followed by numerical simulations accounting for the motion of individual stars orbiting Sgr A*, which allowed the recent history of the accretion rate to be estimated (Cuadra et al. 2006). These estimates ranged typically between $10^{-6}$ to $10^{-5}$ $M_\odot yr^{-1}$. More recently, it was discovered that most of the massive stars are confined to one or two stellar disks (e.g., Paumard et al. 2006). Numerical simulations by Cuadra et al. (2008) accounted for the range of eccentricities of stars in the cluster and found that the mass accretion rate is similar to Bondi rate. However, accurate 3D positions of individual stellar sources is lacking, and knowledge of the loss of angular momentum of gas very close to Sgr A* is poor due to the limited resolution of the simulations. It turns out that the mass accretion rate depends critically on the net angular momentum of ionizing gas in the strong gravitational potential of Sgr A*. High angular momentum of stellar winds near Sgr A* can reduce the mass accretion rate substantially. Furthermore, the accretion rate also depends on the assumed homogeneity of the medium. This may not be correct in the region within a few arcseconds of Sgr A*, given that infalling, dense neutral gas may be present from in-situ star formation or from infalling clouds such as those of the ionized streamers (i,e., Sgr A West). While there is an estimate of the Bondi accretion rate based on X-ray measurements of Sgr A* (Baganoff 2003), it is not clear what the contribution to the X-ray flux is from young stellar sources in the inner one arcsecond of Sgr A*. The steady source of X-ray emission is likely to be an upper limit due to confusing X-ray sources that may be distributed in the inner $1''$ of Sgr A*.

## Radiative Efficiency of the Flow

The second parameter that may be important in lowering the accretion luminosity of Sgr A* is the radiative efficiency of the accreting material in the inner disk. A variety of radiatively inefficient models have been proposed to explain the steady, broad band emission from Sgr A* by fitting its spectral energy distribution (SED) – for example a thin accretion disk, a disk and jet, outflow, an advection-dominated accretion flow, radiatively inefficient accretion flow, and advection-dominated inflow-outflow solutions (Yusef-Zadeh et al. 2009 and references cited therein). Most of these studies have concentrated on the global structure of the flow in the disk of Sgr A*.

To place a better constraint on the physical characteristics of the flow locally, we have been investigating the time variability of the emission from Sgr A*. Several multi-wavelength studies of Sgr A* with a suite of instruments in different wavelength bands have been carried out in the last several years. Radio studies of the variable emission from Sgr A* have been particularly important as it appears to support that some of the gaseous material expands away from the disk of Sgr A*. The expanding flow can either be bound or unbound to the gravitational potential of Sgr A*. If unbound, it implies that inefficient radiation may in part be due to outflowing material from Sgr A*.

*Expansion of Hot Plasma*

Previous observing campaigns to monitor Sgr A* have found evidence of time delays between the peaking of emission at 43 and 22 GHz. Typical time lags at these frequencies ranges between 20 and 40 minutes, far shorter than expected due to synchrotron losses at these frequencies. The time delay is consistent with a picture in which the synchrotron emission at these wavelengths is initially optically thick. The intensity grows as the blob expands, then peaks and declines at each frequency once the blob becomes optically thin. This first occurs at 43 GHz, and then at 22 GHz about 30 minute later. Time delays ranging up to few hours have also been detected between near-IR, submm and radio frequencies (e.g., Yusef-Zadeh et al. 2009 and references cited therein). These measurements are consistent with the expanding hot plasma blob picture. Modeling of the light curves of Sgr A* in the context of adiabatic expansion of hot plasma provides physical parameters of the emitting region. One parameter that has been estimated in a number of measurements is the speed of the expanding blob being non-relativistic, thus the expanding flow does not escape. However, if the expanding hot plasma blobs show a relativistic bulk motion, as Maitra et al. (2009) have recently argued, then an outflow or a jet is expected to arise from Sgr A*. Additional indirect evidence for an outflow may come from minute time scale variability that we have recently detected. The analysis of optically thick, minute time scale variable emission at radio wavelength implies a brightness temperature of plasma with particle energies that are relativistic. These particles can escape the gravitational potential of Sgr A* in form of outflow. The evidence for outflow from Sgr A* is significant as it can act as a feedback in reducing the partial accretion of ionized stellar winds onto Sgr A*. A more detailed account of thees measurements will be given elsewhere.

## FLARE EMISSION MECHANISM

The near-IR flare emission from Sgr A* has been shown to be highly polarized (Eckart et al. 2006) and is therefore thought to be produced by synchrotron emission from GeV electrons in the $\sim 10$ G magnetic fields. Simultaneous monitoring of Sgr A* in X-rays has shown that X-ray flares are always accompanied by flaring in the IR but that the reverse is not necessarily true. This suggests that the X-ray emission is directly connected with the acceleration of the GeV electrons responsibly for the near-IR synchrotron emission, either through 1) upscattering of submm seed photons by the transient population of GeV electrons responsible for the IR flare emission, 2) upscattering of near-IR flare photons by the electron population responsible for the quasi-steady submm emission, 3) synchrotron self-Compton emission by the transient electron population, or 4) syn-chrotron emission from the high energy tail of the accelerated electron population (see Yusef-Zadeh et al. 2009; Dodds-Eden et al. 2009 and references therein).

Here we focus on the first two inverse Compton scenarios. In the optically thin limit similar spectra and fluxes are produced by upscattering of near-IR flare photons by the quasi-steady electron population responsible for the submm emission and by the upscattering of the submm photons by the transient population of electrons accelerated during the IR flares. However, this degeneracy is broken in Sgr A*, because its SED peaks at submm wavelength implying that the quasi-steady electron population is optically thick to photons below 1000 GHz and optically thin at higher frequencies. Thus the submm seed photons are not efficiently upscattered, whereas the IR seed photons may penetrate the entire population of submm-emitting electrons. As a result, the inverse Compton luminosity produced through the latter scenario dominates that produced by the former scenario. This picture appears to explain quantitatively the observed flux of X-ray flares in all simultaneous X-ray/near-IR flares (Yusef-Zadeh et al. 2009).

While synchrotron or synchrotron-self-Compton emission may well play a role in producing the X-ray flaring in Sgr A*, the ICS scenario is well constrained: the seed photon flux is determined empirically by the measured SED of Sgr A* at mm to submm wavelengths, while the characteristics of the transient electron population responsible for the upscattering are constrained by synchrotron models for the IR flaring. This, in turn, places sharp constraints on the electron density and temperature profile in the inner accretion flow of Sgr A*: fairly standard choices give X-ray fluxes similar to those observed, so some models may overproduce the X-ray flux (Wardle & Pandey, in preparation).

Another constraint arises because the light crossing time over a Schwarzschild radius is roughly 1 minute, so that plausible electron profiles give significant time delays between the IR and X-ray flares because of the distances travelled by IR seed photons before scattering off the electrons elsewhere in the accretion flow. The lack of measurable time delays already provide constraints on the flatness of the electron density and temperature profiles in the vicinity of the hole. These constraints apply regardless of the actual mechanism(s) responsible for the X-ray flares: if ICS is not contributing significantly to the measured X-ray fluxes then the constraints placed on models of the accretion flow become even more severe.